# On the Superposition and Elastic Recoil of Electromagnetic Waves

Hans G. Schantz

Q-Track Corporation

(Email: h.schantz@q-track.com)

*Abstract*—**Superposition demands that a linear combination of solutions to an electromagnetic problem also be a solution. This paper analyzes some very simple problems – the constructive and destructive interferences of short impulse voltage and current waves along an ideal free-space transmission line. When voltage waves constructively interfere, the superposition has twice the electrical energy of the individual waveforms because current goes to zero, converting magnetic to electrical energy. When voltage waves destructively interfere, the superposition has no electrical energy because it transforms to magnetic energy. Although the impedance of the individual waves is that of free space, a superposition of waves may exhibit arbitrary impedance. Further, interferences of identical waveforms allow no energy transfer between opposite ends of a transmission line. The waves appear to recoil elastically one from another. Although alternate interpretations are possible, these appear less likely. Similar phenomenology arises in waves of arbitrary shape and those propagating in free space as well. We may also interpret this behavior as each wave reflecting from the impedance variations the superposition imparts on free space. This work has practical implications to quantum mechanics, field diversity antenna systems, and near-field electromagnetic ranging.**

## I. INTRODUCTION

Superposition requires that a linear combination of solutions to an electromagnetic problem is also a solution. The principle of superposition applies to voltage and current on transmission lines, to electric and magnetic fields in free space, and to energy in both these contexts.

This paper examines a simple example of an electromagnetic wave: a one-dimensional (1-D) wave comprising voltage and current impulses on an ideal, free-space transmission line. This paper analyzes constructive and destructive interferences between identical and inverted short-duration impulse signals. This analysis demonstrates that the instantaneous impedance for an electromagnetic wave is not fixed at the expected characteristic or intrinsic impedance of $Z_s = \sqrt{\mu_0/\varepsilon_0}$, but may assume any value. Further, power is zero for all time at the location of the interference. This implies that the energy from each incident wave reflects, rebounds, or recoils elastically off the other.

Alternate interpretations are possible, but appear less likely. Although laid out in the context of transmission lines, one may readily extend the conclusions of this paper to electromagnetic waves of arbitrary shape and those in free space. We may also interpret electromagnetic elastic recoil as each wave reflecting from the impedance discontinuity that the superposition imparts on free space. This paper demonstrates that different mathematical models are available to describe electromagnetic behavior. We may choose among models by considering their physical implications. This paper further notes applications of electromagnetic recoil phenomena for quantum mechanics, near-field electromagnetic ranging, and for field-diversity antenna systems. First however, this paper considers the origins of superposition.

## II. DISCOVERY OF THE SUPERPOSITION PRINCIPLE

James Clerk Maxwell (1831-1879) recognized the principle of superposition, saying, "...one electrical phenomenon at least, that called electrification by induction, is such that the effect of the whole electrification is the sum of the effects due to the different parts of the electrification. The different electrical phenomena, however, are so intimately connected with each other that we are led to infer that all other electrical phenomena may be regarded as composed of parts, each part being due to a corresponding part of the electrification" [1].

The principle of superposition predates Maxwell, however. Leonardo da Vinci (1452-1519) may be credited with an early formulation of the principle [2]. A more rigorous definition arose in the study of linear mechanical systems. Daniel Bernoulli (1700-1782) established that the motion of a string comprises a superposition of harmonic vibrations. Leonhard Euler (1707-1783) was skeptical of the harmonic approach and thought it incompatible with the equations for travelling waves he pioneered along with Jean le Rond D'Alembert (1717-1783). The controversy remained until Jean Baptiste Joseph Fourier (1768-1830) demonstrated the equivalence of the harmonic and wave approaches [3].

In his classic work On the Sensations of Tone, Herman von Helmholtz (1821-1894) observed, "This law is of extreme importance in the theory of sound, because it reduces the consideration of compound cases to those of simple ones" [4]. So long as mechanical displacements are small, superposition is a good approximation. As vibrations become so large that the square of the displacement has an appreciable influence, superposition breaks down. We see this in the electromagnetic context as well – extremely high-frequency, high-energy electromagnetic waves can give rise to non-linear effects like the production of an electron-positron pair. For most practical RF applications, however, superposition is a valid supposition.

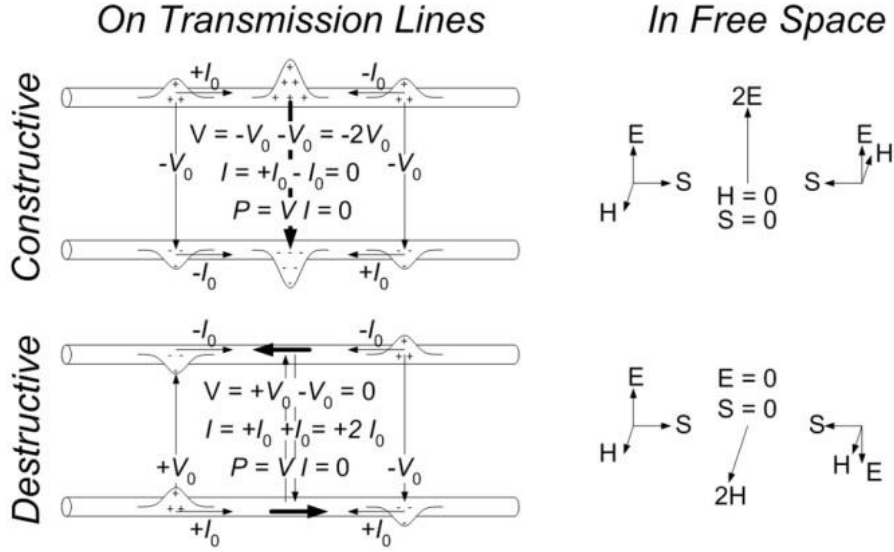

Fig. 1. Summary of constructive and destructive interference on transmission lines and in free space.

### III. SUPERPOSITION AND ENERGY CONSERVATION

At first glance, superposition appears difficult to reconcile with conservation of energy. Assume we have two electric field signals of intensity E at two separate infinitesimal points that combine constructively so that:

$$|\mathbf{E}_{tot}| = |\mathbf{E}_1 + \mathbf{E}_2| = E + E = 2E \tag{1}$$

Before the constructive interference, the electric energy density is:

$$u = \tfrac{1}{2}\varepsilon_0|\mathbf{E}_1|^2 + \tfrac{1}{2}\varepsilon_0|\mathbf{E}_2|^2 = \varepsilon_0 E^2 \tag{2}$$

At the instant of constructive interference, the electric energy density is:

$$u = \tfrac{1}{2}\varepsilon_0|\mathbf{E}_1 + \mathbf{E}_2|^2 = \tfrac{1}{2}\varepsilon_0\left(|\mathbf{E}_1|^2 + 2|\mathbf{E}_1 \cdot \mathbf{E}_2| + |\mathbf{E}_2|^2\right)$$
$$= \tfrac{1}{2}\varepsilon_0\left(E^2 + 2E^2 + E^2\right) = 2\varepsilon_0 E^2 \tag{3}$$

The electric energy density doubles at the instant of superposition, a result difficult to reconcile with conservation of energy.

Destructive interference poses a similar challenge. Destructive interference between our two infinitesimal signals requires:

$$|\mathbf{E}_{tot}| = |\mathbf{E}_1 + \mathbf{E}_2| = E - E = 0 \tag{4}$$

Thus, at the instant of destructive interference, the electric energy density is:

$$u = \tfrac{1}{2}\varepsilon_0|\mathbf{E}_1 + \mathbf{E}_2|^2 = \tfrac{1}{2}\varepsilon_0\left(|\mathbf{E}_1|^2 + 2|\mathbf{E}_1 \cdot \mathbf{E}_2| + |\mathbf{E}_2|^2\right)$$
$$= \tfrac{1}{2}\varepsilon_0\left(E^2 - 2E^2 + E^2\right) = \tfrac{1}{2}\varepsilon_0(0)^2 = 0 \tag{5}$$

The electric energy density vanishes at the instant of superposition, a result also difficult to reconcile with conservation of energy.

These apparent paradoxes are resolved upon examination of the larger context as illustrated in Figure 1. Two electromagnetic waves propagating in opposing directions with aligned electric fields will have opposed magnetic fields. Thus a constructive interference of the electric component of the waves requires a destructive interference of the magnetic component and vice versa. The electric energy in a constructive interference is indeed twice the electric energy of the original waves, because all the original magnetic field energy transforms to electric energy. Similarly, the electric energy vanishes in a destructive interference because all the electric energy has become magnetic energy. By this simple yet elegant interaction between electric and magnetic energy, nature allows both superposition and conservation of energy to be upheld. A similar interplay between electric and magnetic energy occurs in the superposition of voltage and current waves on a transmission line, as depicted in Figure 1.

As two waves interfere and superimpose, they exhibit resonance-like behavior where the instantaneous impedance deviates from intrinsic or characteristic free-space value ($Z_S$) and becomes either arbitrarily large (as magnetic field or current go to zero) or arbitrarily small (as electric field or voltage goes to zero). Despite the simplicity of this result, there is little discussion of this phenomena in the literature. The only comprehensive discussion of the topic known to the author is due to W.S. Franklin in 1909 [5]. Franklin distinguishes between "pure" waves with equal portions of electric and magnetic energy and "impure" waves in which the balance has been disrupted to favor either electric or magnetic energy. An impure wave results from and gives rise to two oppositely travelling pure waves. In more recent engineering texts, the author found only a single reference to the deviation of wave impedance from characteristic impedance in the presence of a reflected wave [6] and others have commented that "discussion of this is sparse in the textbooks" [7].

Although the results depicted in Figure 1 illustrate the way in which interfering waves satisfy both superposition and energy conservation, the results raise further troubling questions. At the moment of superposition, either the electric or magnetic field goes to zero, so $\mathbf{S} = \mathbf{E} \times \mathbf{H}$ (the Poynting Vector) goes to zero. The resulting all-magnetic or all-electric field may be thought of as momentarily static because at that instant it is at rest with no flow of energy. Yet electromagnetic waves propagate at the speed of light – how can they "slow down" to become momentarily static and then continue on past the location of the interference as if nothing happened? We can take a closer look at this behavior by considering how one-dimensional electromagnetic waves propagate on transmission lines.

## IV. Superposition of Mirror-Image Waves

This section considers the behavior of voltage and current waves along an ideal transmission line in free space. Thus, the line is characterized by permittivity $\varepsilon_0$ and permeability $\mu_0$. Consider two symmetric voltage waves on this line. One propagates forward, the other in reverse, so that they superimpose constructively at $(z, t) = (0, 0)$:

$$V(z,t) = V_+(z,t) + V_-(z,t)$$
$$= V_0 F_+(ct-z) + V_0 F_-(ct+z) \quad (6)$$

where $z$ is the coordinate along the line, $t$ is time, $c$ is the speed of light, and $V_0$ is a constant with units of voltage per length. The actual space-time dependence "$F$" is completely arbitrary except that we assume finite or compact, non-overlapping wave packets, and we assume that the forward propagating wave is a spatially reversed or mirror-image version of the reverse propagating wave. In other words: $F_+(z) = F_-(-z) = F(z)$. Note that the speed of propagation is the speed of light:

$$c = \frac{1}{\sqrt{LC}} = \frac{1}{\sqrt{\varepsilon_0 \mu_0}} \quad (7)$$

where $L$ is the inductance per unit length and $C$ is the capacitance per unit length. This voltage wave has a corresponding current wave:

$$I(z,t) = I_+(z,t) + I_-(z,t) = I_0 F_+(ct-z) - I_0 F_-(ct+z) \quad (8)$$

where $I_0$ is a constant with units of current per length. This section combines an analytic and a graphical space-time presentation to constructive and destructive interference between the waveforms. The graphical waveforms are the third time derivative of a Gaussian function. An appendix provides Mathematica code for generating the space-time diagrams, in case the mathematical details are of interest. The mathematical analysis applies to arbitrary "mirror-image" waveforms as defined above.

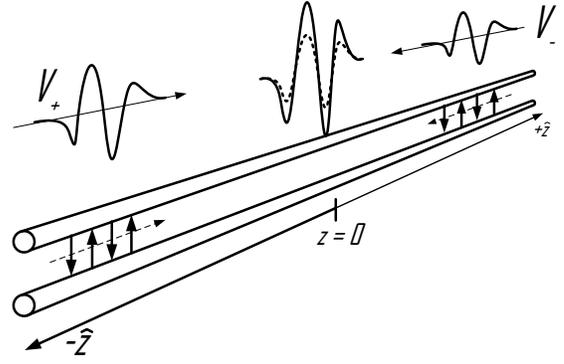

Fig. 2. In constructive interference, voltage is double that of an individual waveform, so the total electric energy is four times that of an individual wave. Total electric energy doubles for the duration of the superposition.

### A. Constructive Interference

Consider identical voltage waves with identical relative time dependence. The waves propagate in opposite directions along an ideal transmission line, so as to superimpose precisely at $t = 0$. At that instant there will be a waveform with the same time dependence as either of the individual waveforms, but with twice the voltage amplitude:

$$V(z,t=0) = V_0 F_+(ct-z) + V_0 F_-(ct+z) = 2V_0 F(z) \quad (9)$$

This is a case of "constructive interference." However, since the electric energy is proportional to the voltage squared, the total electric energy is four times that of an individual wave. The electric energy density under condition of the constructive interference at $t = 0$ is:

$$u_E(z,t=0) = \frac{1}{2} CV_0^2 \left(F_+^2(-z) + 2F_+(-z)F_-(z) + 2F_-^2(+z)\right)$$
$$= \frac{1}{2} CV_0^2 F^2(z) + CV_0^2 F^2(z) + \frac{1}{2} CV_0^2 F^2(z) \quad (10)$$
$$= 4\left(\frac{1}{2} CV_0^2 F^2(z)\right)$$

With respect to electric energy, we have a case of $1 + 1 = 4$. Fig. 2 shows this constructive interference.

Note that a **constructive** interference of voltage waves necessarily requires a **destructive** interference of their corresponding current waves (and vice versa).

$$I(z,t) = I_0 F_+(-z) - I_0 F_-(+z) = I_0 F(z) - I_0 F(z) = 0 \quad (11)$$

Total energy is conserved, however, because the magnetic energy is zero:

$$u_H(z,t=0) = \frac{1}{2} LI_0^2 F_+^2(-z) - LI_0^2 F_+(-z)F_-(+z)$$
$$+ \frac{1}{2} LI_0^2 F_-^2(+z) \quad (12)$$
$$= \frac{1}{2} LI_0^2 F^2(z) - LI_0^2 F^2(z) + \frac{1}{2} LI_0^2 F^2(z) = 0$$

The excess electric energy is due to a transformation of magnetic energy. Momentarily, the equal mix of electric and magnetic energy that characterizes an electromagnetic wave transforms entirely to electrostatic energy.

An electromagnetic wave in free space propagates at the speed of light as does its associated energy. Yet in this case of interference, energy comes to a momentary rest and then appears to continue its progress as if it had been travelling at the speed of light all along. This confusing behavior becomes clearer on considering the power:

$$P(z,t) = V(z,t)I(z,t) = \left(V_0 F_+(ct-z) + V_0 F_-(ct+z)\right)$$
$$\times \left(I_0 F_+(ct-z) - I_0 F_-(ct+z)\right) \qquad (13)$$
$$= V_0 I_0 F_+^2(ct-z) - V_0 I_0 F_-^2(ct+z)$$

which divides neatly into a positive forward propagating power and a negative reverse propagating power. Of course, at $t = 0$, the power is instantaneously equal to zero everywhere because the current goes to zero. Another way to look at it is that the forward and reverse propagating power exactly cancel everywhere at $t = 0$:

$$P(z, t=0) = V_0 I_0 F_+^2(-z) - V_0 I_0 F_-^2(+z)$$
$$= V_0 I_0 F^2(z) - V_0 I_0 F^2(z) = 0 \qquad (14)$$

This supports the observation above that the "propagating" electromagnetic energy is instantaneously static at $t = 0$ for all $z$. By the symmetry of the waveforms, however, since $F_+(z) = F_-(z) = F(z)$, it necessarily follows that $F_+(ct) = F_-(ct) = F(ct)$. Thus, the power has to always be exactly zero at $z = 0$ for all time as well:

$$P(z=0,t) = V_0 I_0 F_+^2(ct) - V_0 I_0 F_-^2(ct)$$
$$= V_0 I_0 F^2(ct) - V_0 I_0 F^2(ct) = 0 \qquad (15)$$

This result leads to a remarkable conclusion. No energy transfers between the $-z$ and $+z$ halves of the transmission line. The $-z$ energy and the $+z$ energy remain partitioned each on their own side. The forward and reverse propagating waves elastically recoil, or bounce off each other. The following section further employs space-time diagrams to further examine this behavior for specific waveforms.

*B. Space-Time Analysis*

Further insight follows from the space time diagrams of Fig. 3. Each plot depicts time ($t$) on the vertical axis (scaled by "$c$") and location ($z$) on the horizontal axis. Fig. 3(a) shows the constructive interference of voltage signals $V(z, t) = V_+(z, t) + V_-(z, t)$. Observe how $V(z, t) = 0$ at $z = 0$ for all time. Fig. 3(b) presents a space time diagram of electric energy. Note the twin peaks in electric energy at $t = 0$ as the voltage signals add together constructively. Instantaneously, at $t = 0$ the total electric energy doubles.

The mystery of how two waveforms with unit electric energy can combine to yield four times their individual unit electric energies becomes clear upon examining the behavior of the current shown in the space time diagram of Fig. 3(c). While the voltage adds up constructively, the current $I(z, t) = I_+(z, t) + I_-(z, t)$ adds up destructively. Observe that $I(z, t) = 0$ at $t = 0$. Thus, the magnetic energy, portrayed in the space-time diagram of Fig. 3(d), goes to zero at $t = 0$. The magnetic energy in each wave transforms to electric energy for the duration of the overlap.

Additional clues are evident in Fig. 3(e), the space-time diagram for the power – the product of the voltage and the current. The wave travelling in the positive direction exhibits positive power. The wave travelling in the negative direction displays a negative power. In the region of the overlap, the power goes to zero, not only for $t = 0$ but also for $z = 0$. As previously noted, this means that at $t = 0$, the system of two propagating electromagnetic waves is instantaneously static, with no flow of energy.

The space-time diagram for the power of Fig. 3(e) demonstrates there is no energy transfer between the $-z$ and $+z$ halves of the transmission line. Instead, there is a resonance-like behavior as energy transitions from the equal mix of electric and magnetic energy characteristic of a typical electromagnetic wave into all-magnetic, then all-electric, then all-magnetic before rebounding, decoupling and resuming the balanced mix of a typical wave.

Fig. 3(f) presents a space-time diagram for the energy. The energy trajectories cross showing the superposition of the energy in four prominent peaks. Comparison of the energy peaks to the power space-time diagram of Fig. 3(e) shows the energy peaks coincide with space-time locations where and when the power is zero. Thus the energy peaks denote locations where the forward travelling and reverse propagating energy rebound and reflect from each other. The energy peaks on the $t = 0$ axis represent momentarily static electric energy, and the energy peaks on the $z = 0$ axis result from instantaneously static magnetic energy.

In summary, the total electric energy is indeed four times that of an individual wave, however, the total electromagnetic energy is conserved. The excess electric energy arises from the conversion of magnetic to electric energy. As the waves interfere and superimpose, they exhibit resonance-like behavior where the instantaneous impedance deviates from intrinsic or characteristic free-space value ($Z_S$) and becomes either arbitrarily large (as current goes to zero) or arbitrarily small (as voltage goes to zero) [6]. In addition, the impedance changes sign depending upon whether energy propagates in the forward or reverse direction.

*C. Impedance*

Impedance is a generalization of the concept of resistance, so a brief discussion of resistance would be in order. "Resistance" has a dual meaning. Resistance is the ratio of voltage to current in a simple electrical circuit: $R = V/I$. Resistance is also an intrinsic property of certain circuit elements (i.e., "resistors") that give rise to the corresponding ratio of voltage to current. A 1kohm resistor gives rise to a 1000:1 ratio between the electric potential (in volts) and the current (in amps). In the context of an alternating current (AC) circuit, however, the dual meanings of resistance diverge. The actual ratio of voltage to current depends upon the reactive character of the resistor – extra capacitance or inductance that store electric and/or magnetic energy. Thus the concept of resistance had to be generalized to embrace the broader context of time-varying or AC electrical systems.

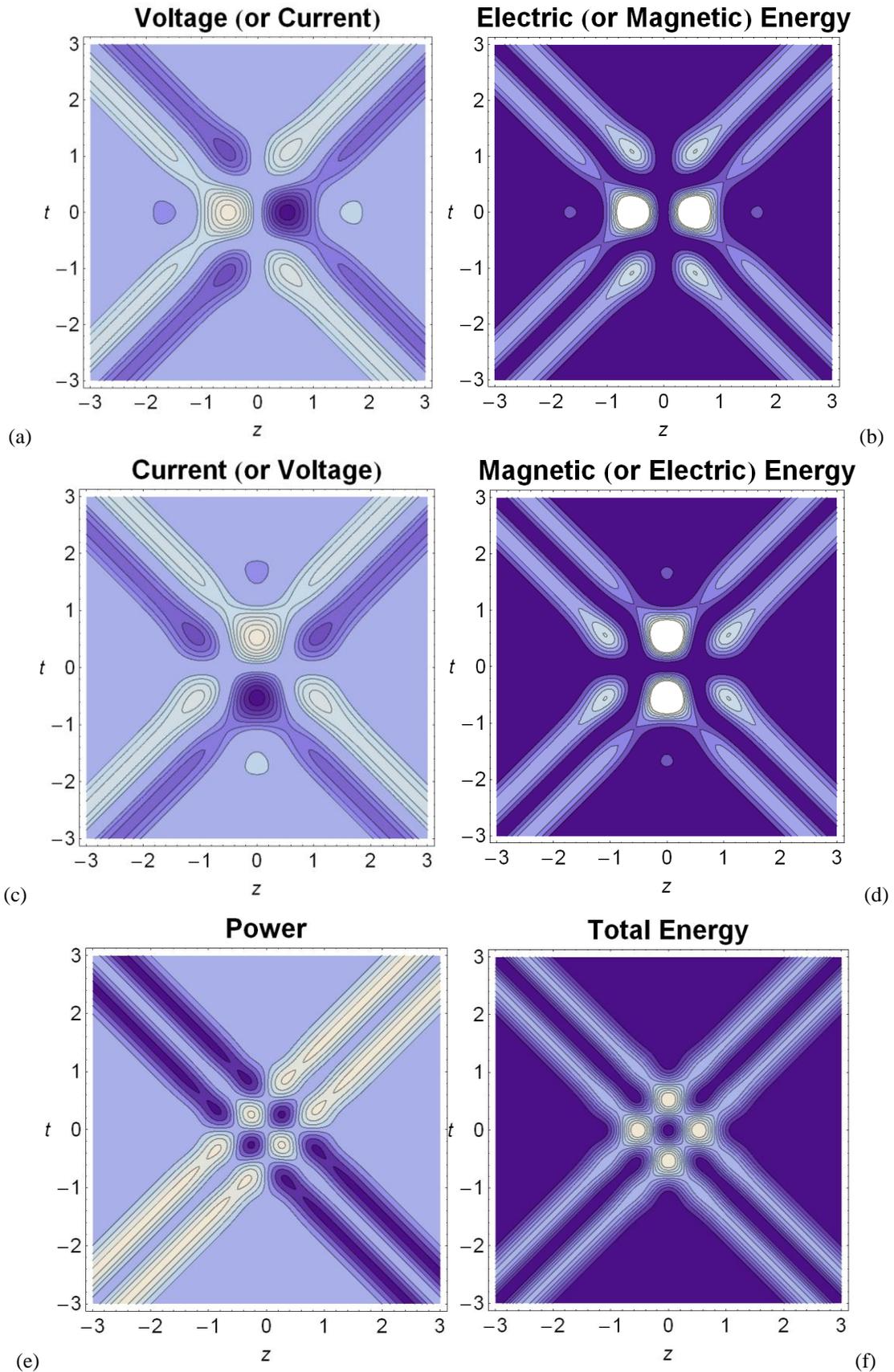

**Fig 3.** Space-time diagrams of constructive (or destructive) interference between two waves. On diagrams to the left (a, c, e) light indicates positive values, dark negative. Diagrams to the right (b, d, f) are positive definite so black denotes zero.

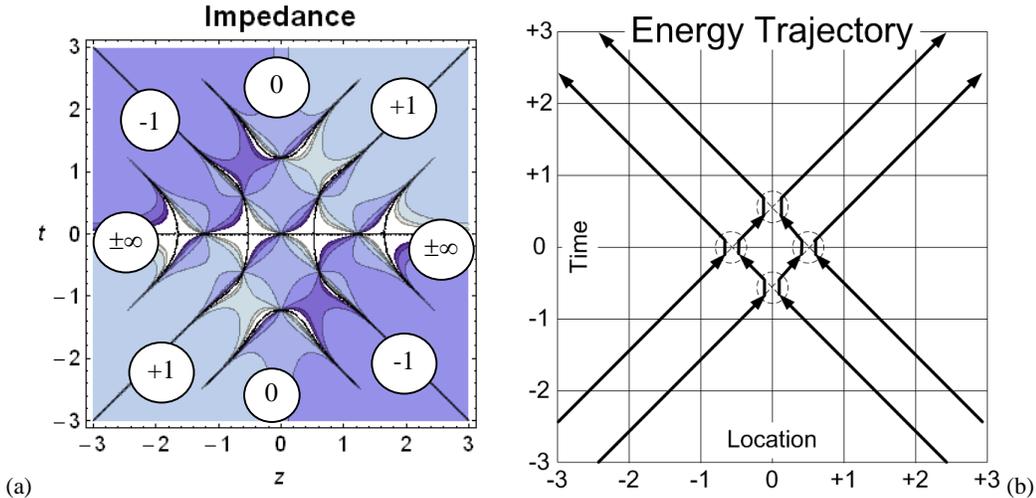

**Fig. 4.** (a) The instantaneous impedance of two electromagnetic waves deviates significantly from the characteristic value under interference between two waves, and (b) a qualitative sketch of the energy trajectory for the principal lobes of the waveforms.

Sergei Schelkunoff (1897-1992) further generalized the concept of impedance to encompass electromagnetic systems in free space [8]. Schelkunoff defined impedance as $Z = E/H$ where the particular choice of field components depends upon the geometry. While noting that the intrinsic impedance of free space is $Z_s = \sqrt{\mu_0/\varepsilon_0} = 376.6\Omega$, Schelkunoff described how the actual ratio of electric to magnetic field deviates from the intrinsic value in the close vicinity of a dipole source. Here again, while the typical ratio of electric to magnetic field intensity for a wave in free space is the same as the characteristic or intrinsic impedance, the impedance can be arbitrarily large or small in the close vicinity of an electric or magnetic dipole source, respectively. Similar deviations from the intrinsic impedance of free space occur in conjunction with superposition or in standing waves on transmission lines [6].

Like resistance, impedance has a dual nature. On the one hand impedance describes the intrinsic ratio of potential to current or electric to magnetic field characteristic of "pure" waves propagating though a particular medium. On the other hand impedance also describes the actual instantaneous ratio. Depending on the situation, the actual instantaneous impedance may diverge from the typical intrinsic or characteristic value in a medium.

Finally, Schelkunoff observed that impedance may be taken as a signed quantity thus denoting a propagation of energy in a forward (positive) or reverse (negative) direction. This paper adopts Schelkunoff's formulation to represent impedance as, in effect, a 1-D vector with the sign indicating direction of propagation.

Fig. 4(a) presents a space-time diagram of the instantaneous impedance ($Z = V/I$) normalized to the characteristic impedance $Z_S$. Asymptotically, the forward traveling wave has impedance "+1" (i.e. + $Z_S$) as seen in the lower left and upper right quadrants. The reverse travelling wave has impedance "-1" (i.e. $Z_S$) as seen in the lower right and upper left quadrants. Along the $z = 0$ axis, the voltage goes to zero, so impedance is zero. Along the $t = 0$ axis, current goes to zero, so the impedance becomes either positively infinite (light) or negatively infinite (dark).

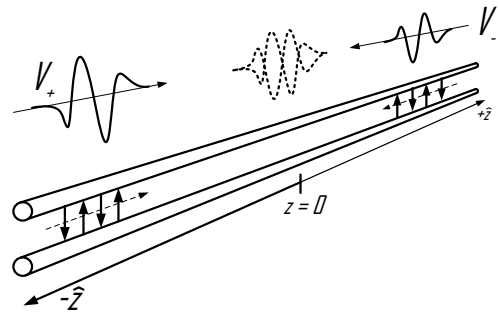

**Fig. 5.** In destructive interference, the voltage is zero, so the total electric energy is zero for the duration of the superposition.

Fig. 4(b) shows a qualitative sketch of the energy trajectories for the two principal lobes of each waveform. The absence of any power at $z = 0$ means no net energy passes from the negative to the positive $z$ side of the transmission line. In fact, as two matched waveforms interfere and rebound each from the other, the reflected leading edge or "head" of each waveform interferes with its own incoming trailing edge or "tail." Upon reflection, the energy in the tail of the forward propagating waveform becomes the energy in the head of the reverse propagating waveform. In other words, the energy in the tail of the waveform never gets closer to $z = 0$ than about half the spatial length of the waveform.

### D. Destructive Interference

Now, consider the same situation but with one of the two waveforms inverted so as to precisely cancel out the other. At the instant the waves overlap the voltage is everywhere zero along the transmission line. Since the voltage is zero, the total electric energy is also zero. Here, we have a case of $1 + 1 = 0$. Fig. 5 shows this destructive interference.

The destructive interference of two voltage waveforms turns out to be the dual of the case of constructive interference. Mathematically, we may swap voltage for current and vice versa in the previous analysis. The space-time diagrams of Fig. 3 describe this situation as well, provided we swap voltage for current and electric energy for magnetic. In fact, an additional fascinating duality is evident. Voltage and current are related by a ninety degree rotation that swaps the time and space axes.

### E.  Arbitrary Waveforms

This paper relies on simple and symmetric "mirror-image" waveforms to illustrate the physics whereby electromagnetic waves recoil, rebound, or bounce off each other. A Newton's Cradle illustrates the conservation and energy and momentum through the interaction of multiple identical balls anchored to pendulums of identical length. The examples of this paper present an electromagnetic Newton's Cradle showing the perfect reflection of electromagnetic waves from each other under ideal conditions to illustrate electromagnetic recoil under the simplest circumstances.

The more general case of interaction between arbitrary electromagnetic waves involves reflection of a portion of the propagating energy. Whenever a standing wave occurs, the actual field impedance deviates from the intrinsic or characteristic impedance of the medium. A portion of the propagating energy momentarily becomes either electrostatic or magnetostatic. Superluminal propagation of energy would be required to catch up with the wave front propagating at the speed of light. Since superluminal propagation contradicts well-established electromagnetic theory, any standing wave represents a collision between electromagnetic waves in which energy recoils, each wave exchanging energy with the other.

Oliver Heaviside (1850-1925) derived the concept of an "energy velocity" for electromagnetic phenomena in the context of plane waves [9]. We can understand the normalized energy velocity "$\gamma = v/c$" of electromagnetic energy in a transmission line by comparing the power (i.e., the dynamic energy per time) to the total energy per unit length $U_L$:

$$\gamma = \frac{v}{c} = \frac{P}{cU_L} = \frac{VI}{c\left(\frac{1}{2}LI^2 + \frac{1}{2}CV^2\right)}$$
$$= \frac{2\frac{V}{I}}{\left(Z_0 + \frac{1}{Z_0}\left(\frac{V}{I}\right)^2\right)} = \frac{2z}{(1+z^2)} \quad (16)$$

The speed of light relates the time associated with the power to the length associated with the energy per unit length. The normalization to the speed of light allows the equation to be expressed in terms of normalized impedance. Normalized impedance is the actual impedance $Z = V/I$ normalized to the intrinsic or characteristic impedance $Z_S$ to yield a dimensionless quantity $z = Z/Z_S$. In the context of free space fields instead of transmission lines, we may similarly define the same ratio:

$$\gamma = \frac{v}{c} = \frac{S}{cu} = \frac{|\mathbf{E} \times \mathbf{H}|}{c\left(\frac{1}{2}\varepsilon_0|\mathbf{E}|^2 + \frac{1}{2}\mu_0|\mathbf{H}|^2\right)}$$
$$= \frac{2EH}{\sqrt{\frac{\varepsilon_0}{\mu_0}}|E|^2 + \sqrt{\frac{\mu_0}{\varepsilon_0}}|H|^2} = \frac{2\frac{E}{H}}{\frac{1}{Z_S}\left(\frac{E}{H}\right)^2 + Z_S} = \frac{2z}{(1+z^2)} \quad (17)$$

where we assume that $\mathbf{E} \perp \mathbf{H}$ so $|\mathbf{E} \times \mathbf{H}| = EH$. When $\gamma = 1$, all the energy at a point is dynamic, and therefore freely propagating. When $\gamma = 0$, all the energy at a point is static, and therefore involved in an elastic collision.

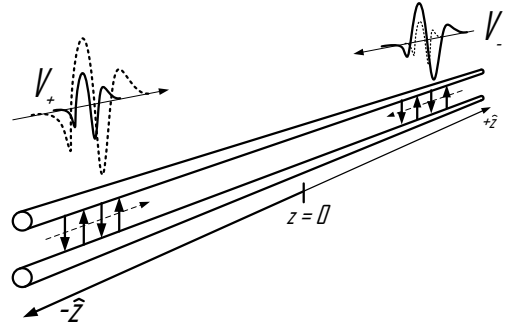

**Fig. 6.** A smaller time compressed waveform interacts with a larger waveform.

Interestingly, this is very similar to the result from microwave theory for the transmitted fraction of incident power at an impedance discontinuity:

$$\frac{P_{TX}}{P_{inc}} = 1 - |\Gamma|^2 = 1 - \left(\frac{z-1}{z+1}\right)^2 = \frac{4z}{(1+z)^2}$$
$$= \frac{\frac{1}{c}|\mathbf{E} \times \mathbf{H}|}{\frac{1}{4}\varepsilon_0|\mathbf{E}|^2 + \frac{1}{2c}|\mathbf{E} \times \mathbf{H}| + \frac{1}{4}\mu_0|\mathbf{H}|^2} \quad (18)$$

where $\Gamma$ is the voltage reflection coefficient. At $z = 1$, (17) and (18) agree. As $z \to 0$ or as $z \to \infty$, there is a factor of two difference. In a sense, we may think of the interaction between electromagnetic waves as each reflecting off the impedance discontinuities imparted on free space by the other. The analogy is not exact, however. In the microwave circuit case, impedance is a fixed intrinsic property of the medium and the transmission line geometry. In electromagnetic interactions, impedance is a dynamic quantity governed by the instantaneous superposition of the electromagnetic waves. Thus, we should not be surprised that the physics is not exactly the same.

The particulars of the interaction between arbitrary waveforms depend upon the details of the waveforms. If one is larger than the other, the smaller imparts its energy to the leading edge of the larger, and then the larger provides the energy for the smaller to continue on its own way. Consider two waveforms similar to those of the previous section, except that one is reduced in amplitude and compressed in time duration relative to the other. Fig. 6 shows this situation.

Energy velocity provides a mathematical tool for analyzing the flow of energy in interactions of this kind. Fig. 7a presents a space-time diagram with a line contour plot of the energy density along with arrows showing the instantaneous energy velocity as the two waveforms depicted in Fig. 6 interact. Energy from the smaller waveform accretes to the leading edge of the larger. The trailing edge of the larger waveform then supplies the energy to comprise the smaller waveform. Fig. 7b presents a sketch of the energy worldlines. A detailed space-time examination of energy velocity helps reveal the specifics of the interaction.

**Total Energy**

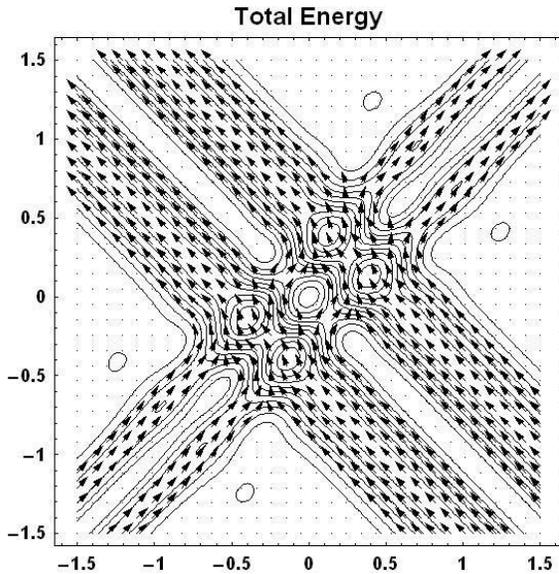
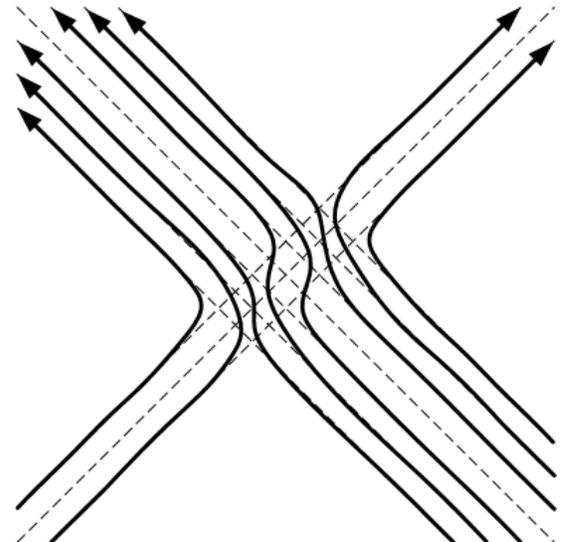

**Fig. 7a (left).** When arbitrary waveforms interact, the energy from the smaller waveform accretes to the leading edge of the larger. The larger waveform sheds enough energy from its trailing edge to comprise the smaller waveform. Contours show local energy density and arrows denote energy velocity.

**Fig. 7b (right).** The energy flow pattern of Figure 6a may be depicted by sketching the energy worldlines.

## V. INTERPRETATION AND IMPLICATIONS

In the case of constructive interference between voltage waveforms, the voltage doubles, but the current, and thus the magnetic energy, go to zero. The total electric energy is four times that of an individual wave or twice that of the total starting electric energy, but the apparent excess of energy has come from the conversion of magnetic to electric energy. In the case of destructive interference between voltage waveforms, the voltage cancels, but the current doubles, and thus the total magnetic energy increases fourfold. Now, the total magnetic energy is four times that of an individual wave or twice that of the total starting magnetic energy. The apparently vanishing electric energy has merely transformed to magnetic energy.

When propagating electromagnetic waves interfere, they give rise to momentarily static concentrations of electric or magnetic energy. Power is zero at the point of interference because current is zero (in the case of constructive interference) or because voltage is zero (in the case of destructive interference). Because power is zero, there is no net energy transfer at the point of interference. To the extent we can associate electromagnetic waves with the energy they carry, the waves appears to recoil elastically in the case of a perfect constructive or destructive interference between identical waves. In the more general case, the forward and reverse propagating waves exchange energy. The energy in the trailing edge of the forward propagating wave becomes the energy in the leading edge of the reverse propagating wave. The energy in the trailing edge of the reverse propagating wave becomes the energy in the leading edge of the forward propagating wave. This section discusses a mechanical analogy and presents a variety of alternate interpretations for the results of these interference examples.

### A. A Mechanical Analogy

Interestingly, similar interference phenomena occur in the case of longitudinal compression waves in a uniform bar [10]. If identical compression waves propagating in opposite directions overlap and interfere, there will be doubled stress and zero velocity in the region of superposition. Since velocity is zero, the boundary condition in the overlap region of the bar is identical to fixing the end of the bar. If the compression waves are equal and opposite, when they overlap the velocity doubles and the pressure goes to zero. This boundary condition is equivalent to a free end of the bar. In both cases, the physics is similar – total energy is conserved. Energy merely transforms between potential energy (electric or compressional) and kinetic energy (magnetic or kinetic) [7].

### B. Alternate Physical Interpretations

What is a wave? Some, like Heaviside, argued that fields are fundamental and potentials secondary [11]. Others cite the Aharonov-Bohm effect to argue for the primacy of potentials [12]. Still others appeal to Einstein's mass-energy equivalence to reify energy as a fundamental substance instead of being a characteristic or property of physical systems [13]. Resolving these mutually contradictory points of view lies beyond the scope of this paper. Whatever we imagine electromagnetic waves to be, they have associated with them a particular quantity of energy. By tracking the flow or motion of this energy we gain insight to the behavior of the waves. This paper argues that the energy, and thus the waves, rebound elastically from the location of a perfect constructive or destructive interference. There are a variety of possible alternate explanations for the phenomenology of interference exhibited in these examples, however.

One might argue that the waves actually do permeate through each other. The fact that power is zero at the point of interference ($z = 0$) may be interpreted as equal and opposite flows of energy cancelling out to yield zero net power.

However, this is not as simple as two equal and opposite energy flows. If the forward fields ($\mathbf{E}_+$, $\mathbf{H}_+$) and the reverse fields ($\mathbf{E}_-$, $\mathbf{H}_-$) retain their independent identities on superposition, then there are actually four separate power flows, not two:

$$\begin{aligned}
\mathbf{S} &= \left(\mathbf{E}_+ + \mathbf{E}_-\right) \times \left(\mathbf{H}_+ + \mathbf{H}_-\right) \\
&= \left(\mathbf{E}_+ \times \mathbf{H}_+\right) + \left(\mathbf{E}_+ \times \mathbf{H}_-\right) + \left(\mathbf{E}_- \times \mathbf{H}_+\right) + \left(\mathbf{E}_- \times \mathbf{H}_-\right) \quad (19) \\
&\neq \mathbf{S}_+ + \mathbf{S}_-
\end{aligned}$$

This "power permeation" explanation requires us to assume that the multiple sources or causes of an electric phenomenon somehow retain their independent identity upon superposition. Instead of one electric field, this hypothesis requires there to be as many distinct and separate electric fields as there are sources. The principle of parsimony (i.e. Ockham's Razor) suggests this more complicated interpretation is less likely.

The case against power permeation becomes even stronger upon realizing that constructive interference satisfies the exact same boundary conditions as an open at $z = 0$: current is zero. Similarly, destructive interference satisfies the exact same boundary conditions as a short at $z = 0$: voltage is zero. These examples are inverse applications of image theory. In image theory, we replace a reflecting boundary with a mirror "image" of the source. Then we mathematically attribute the reflected signal to the action of the image. In the present transmission line interference examples, we may satisfy the same boundary conditions by introducing an appropriate termination at $z = 0$. The opposing wave creates a virtual electromagnetic termination reflecting the other wave in a way exactly equivalent to that of a real short or open.

In the case of a short or an open, energy penetration through the termination to the other side of the line is difficult to justify [14]. The boundary conditions and the solutions for the voltage and current for terminations are exactly identical to those of the appropriate interference. To justify the power permeation hypothesis, one must offer an explanation for why the identical mathematical descriptions for interference and reflection should have two fundamentally different physical interpretations. The power permeation hypothesis assumes identical mathematical descriptions represent reflection (in the case of termination) and permeation (in the case of interference).

As Isaac Newton observed, "…to the same natural effects we must, as far as possible, assign the same causes" [15]. The more reasonable interpretation is that the electromagnetic energy behaves the same for the case of interference at $z = 0$ as it does for termination at $z = 0$. In each case, instantaneous power is zero at $z = 0$. The energy and thus the waves reflect or recoil.

Finally, we have the observation that energy becomes momentarily static in a superposition, and yet somehow catches up with the respective wavefronts that have continued to move on past the superposition at the speed of light. The advocate of energy permeation requires either some form of superluminal energy transport or a different theory of electromagnetic energy transport than that of Poynting and Heaviside.

## C. Are Physical Interpretations Relevant?

A still further alternate point-of-view holds that physical interpretation is irrelevant. We have a mathematical description of electromagnetic phenomena that describes wave propagation and matches our measurements of the system. Asking what really goes on is irrelevant, according to this viewpoint. This approach begs the question of which mathematical model will prove most useful.

One might employ a number of different mathematical models to describe the phenomenology of wave propagation. For instance, consider the superposition of identical sinusoidal waves propagating in opposite directions:

$$\begin{aligned}
V(z,t) &= V_+(z,t) + V_-(z,t) \\
&= V_0 \sin(\omega t - kz) + V_0 \sin(\omega t + kz) \\
I(z,t) &= I_+(z,t) + I_-(z,t) \\
&= I_0 \sin(\omega t - kz) + I_0 \sin(\omega t + kz)
\end{aligned} \quad \text{(19a, b)}$$

where $k = 2\pi/\lambda$ and $\omega = kc = 2\pi f$. These relations invoke the D'Alembert-Euler travelling-wave approach. Alternatively, one could create an exactly equivalent mathematical description of the phenomenology employing the harmonic formulation pioneered by Daniel Bernoulli:

$$\begin{aligned}
V(z,t) &= V_0 \sin(\omega t - kz) + V_0 \sin(\omega t + kz) \\
&= V_0 \left[\begin{array}{l} \sin \omega t \cos kz - \cos \omega t \sin kz \\ + \sin \omega t \cos kz + \cos \omega t \sin kz \end{array}\right] \quad \text{(20a)} \\
&= 2V_0 \sin \omega t \cos kz
\end{aligned}$$

$$\begin{aligned}
I(z,t) &= I_0 \sin(\omega t - kz) - I_0 \sin(\omega t + kz) \\
&= I_0 \left[\begin{array}{l} \sin \omega t \cos kz - \cos \omega t \sin kz \\ - \sin \omega t \cos kz - \cos \omega t \sin kz \end{array}\right] \quad \text{(20b)} \\
&= -2I_0 \cos \omega t \sin kz
\end{aligned}$$

The equivalence in this special case is obvious. Fourier's demonstration that the two approaches are equivalent in general is sufficiently difficult, however that a mathematical genius like Euler went to his grave convinced the formulations were in fact not equivalent. Which of these mathematical models is "correct?" They both are, mathematically. But by adopting a time domain approach to understanding the energy flow in a standing wave, Kaiser demonstrated that energy oscillates back and forth between successive nodes in a standing wave, recoiling from the resulting superpositions [16].

For another example, one could develop a fully consistent mathematical description of these interference examples under the assumption that the V− wave propagates backward in time, reflects from V+ wave propagating forwards in time, and the interaction causes both waves to reverse their temporal directions of propagation. Such a theory would also be mathematically correct, and of equivalent complexity to the conventional temporal interpretation of wave propagation [17]. A physical interpretation of this theory consistent with the demands of causality would be challenging, however.

Mathematical theories are plentiful. Physical insight is in short supply. Those mathematical models that most closely reflect and describe the underlying physical phenomena lay a foundation for further discoveries. Initially, the heliocentric or Copernican model of the solar system had little to recommend it over the geocentric or Ptolemaic model. Both models employed complicated systems of epicycles and deferents, superimposing multiple circular motions to match the observed positions of the planets. But because Johannes Kepler (1571-1630) started with a Copernican mindset, he was able to look at the observations of Tycho Brahe (1546-1601) and obtain the critical insight that planetary orbits are elliptical, instead of circular. This laid the foundation for Isaac Newton (1642-1727) to discover the Law of Universal Gravitation. In our examinations of electromagnetic phenomena we should always be alert to how and why electromagnetic systems behave the way they do, seeking the deeper understanding that leads to scientific progress.

### D. Extension to Electromagnetic Waves in Free Space

On further reflection, the reader will note that the phenomenology herein described easily extends to the case of plane electromagnetic waves propagating in free space. The shorts and opens of the ideal transmission line may be replaced by perfect electric and magnetic conducting planes interacting with normally incident plane waves, for instance.

Suppose we have a discone antenna source as in Figure 8a. Suppose we wish to model how it interacts with a perfect electrically conducting (PEC) plane some distance away. The plane will reflect the signals. The energy bounces or recoils off the PEC plane. We desire an accurate mathematical description. Applying image theory, we replace the PEC plane with a virtual inverted source as depicted in Figure 8b. By the symmetry of the arrangement, this virtual inverted source will precisely cancel out the electric field at the location of the plane. Thus, the solution on the left side of the plane is identical to the arrangement of Figure 8a.

Suppose instead of a PEC plane, we actually do have a real inverted source, as shown in Figure 8c. By symmetry, the electric field will go to zero on the plane of symmetry. The mathematical solution in Figure 8c is identical to that of Figure 8a. In Figure 8a, we have no trouble characterizing the signals as bouncing off the PEC plane. Should we not similarly characterize the signals in Figure 8c as bouncing off the plane of symmetry?

To justify the power permeation hypothesis, one must be prepared to explain why the identical mathematical descriptions of Figures 8a and 8c lead to fundamentally different physical interpretations. The more parsimonious explanation is that the identical effects share a common cause – specifically that signals bounce off the plane on which electric field goes to zero, whether that plane is a physical PEC, or a "virtual PEC" caused by an equal and opposite signal propagating in the reverse direction.

A full extension of the concepts in this paper to free space electromagnetic waves would rely on the Heaviside-Poynting theory of electromagnetic energy transfer, however, and might be unnecessarily sidetracked into the subtleties of interpreting the Poynting vector and energy flow. The example of waves on an ideal one-dimensional transmission line provides a simple framework within which the behavior of electromagnetic waves during superposition and

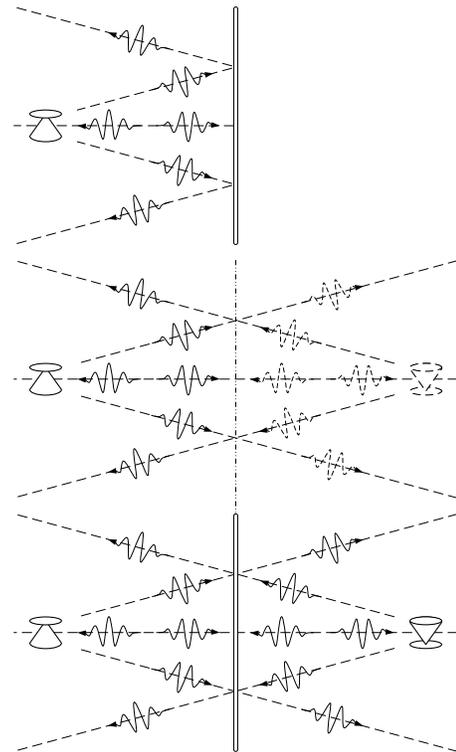

**Fig. 8a (top).** EM signals from a discone antenna are incident on a PEC plane.
**Fig. 8b (center).** By image theory, we may replace the PEC plane with an inverted source.
**Fig. 8c (bottom).** Equal and opposite sources cancel out the E-field on the plane of symmetry. The solution is identical to what we would expect if there were a PEC plane on the plane of symmetry. The identical solutions should have identical interpretations – signals reflect from the plane of symmetry.

interference becomes clear, without any distracting complexity.

The present paper adopts a simplified version of the method of "causal surfaces." This method seeks out locations or surfaces of zero power so as to isolate electromagnetic energy and determine cause and effect relationships. Additional information and examples of this method to waves in free space may be found elsewhere [18-20].

### E. Applications of Electromagnetic Energy Flow

Electromagnetic energy flow streamlines aid in understanding practical EM problems from interference and diffraction [21] to optical phenomena like the spot of Poisson-Arago [22]. This kind of EM flow analysis is also relevant to emerging discoveries in quantum mechanics. The pilot wave perspective of quantum mechanics, pioneered by Louis deBroglie (1892-1987) and David Bohm (1917-1992), treats individual particles or photons as guided by quantum waves along discrete trajectories [23]. For photons, these trajectories are essentially the electromagnetic energy flow streamlines [24]. Recently, Kocsis et al observed average trajectories of single photons in a two-slit interferometer and confirmed that a weak measurement of photons enables the determination of photon trajectories [25]. These quantum mechanical measurements may also be interpreted as measurements of the classical Poynting vector field [26]. The present work suggests that photon-photon interactions are possible at energies and frequencies well below what was previously believed possible.

The lessons of this paper have practical application in understanding applied radio science as well. Radio links often suffer from multiple radio waves taking different paths and combining destructively so as to cancel out the signal making reception difficult. One may mitigate this "multipath" interference in a variety of ways. The present work provides a clear understanding for why "field diversity" offers a similar mitigation to multipath interference. Although one field, say the electric field, may suffer destructive interference from multipath at a point, the other field may yet have useable signal. By implementing a diversity scheme employing both electric and magnetic antennas, one can create a compact receiving array that will be more robust in a multipath environment. An example of such a system has already been implemented and evaluated in the context of short-range near-field wireless links involving electrically-small antennas and links of a half wavelength or less operating at low frequencies [27]. The technique, pioneered by Kwon [28], can be extended to enhance the performance of far-field links as well, in the presence of multipath.

Although typical, line-of-sight, free space waves comprise equal electric and magnetic energy as waves collide and interact the mix may be all electric, all magnetic, or anything in between. This result is also important in near-field electromagnetic ranging [29].

## VI. Conclusion

This paper employed superposition of simple 1-D waves on ideal transmission lines to illustrate some basic and fundamental truths of electromagnetic phenomena. Although we often speak of there being a single, fixed characteristic impedance of free space, as waves superimpose and interfere, actual free space impedance may have an arbitrary value. Superimposed and interfering waves give rise to momentary concentrations of static electric or static magnetic energy associated with the recoil and reflection of interfering waves from each other. In general case of arbitrarily shaped interfering waveforms, the forward and reverse propagating waves exchange energy. The energy in the trailing edge of the forward propagating wave becomes the energy in the leading edge of the reverse propagating wave and vice versa. One interpretation is that each wave reflects off the impedance discontinuity created in free space by the superposition of the two waves.

The ideas in this paper do not represent fundamentally new physics. Rather they provide a different way of looking at how electromagnetics in general and radio waves in particular behave. Physicists and RF engineers refer to "near" fields because their stationary or "reactive" energy will typically be found near to a particular source - within about one wavelength. On the contrary, this paper illustrates how "near" fields are actually all around us. Radio waves interact and combine with each other all the time, generating near fields even arbitrarily far away from the transmitters which create them and the receivers which detect them.

The harmonic approach pioneered by Bernoulli and formalized by Fourier dominates contemporary thought on electromagnetic behavior. This approach tends to deal in time-average characteristics of electromagnetic systems, obscuring or ignoring the detailed time-domain characteristics essential to understanding how electromagnetic systems truly work. The D'Alembert-Euler time-domain wave approach offers remarkable insights to the physics underlying electromagnetic processes.

### Acknowledgements:

The author gratefully acknowledges helpful discussions with Travis Norton, Kirk T. McDonald, Kazimierz Siwiak, Gerald Kaiser, Timothy J. Maloney, Zahra Sotoudeh, and an anonymous reviewer. The author also thanks the organizers of the 2014 Texas Wireless & Microwave Circuits & Systems conference for the opportunity to present an early draft as an invited talk and thanks attendees for their feedback.

### Appendix: Mathematica Code

```
(* Waveforms & Constants: *)
    F:=¹/₈ √(π) exp[−(t − z)²] (12 (t − z) − 8 (t − z)³)
    G:=¹/₈ √(π) exp[−(t + z)²] (12 (t + z) − 8 (t + z)³)
    m:=3 (*Scale min/max *); p:= 200 (* PlotPoints *)
(* Voltage (or Current) *)
    ContourPlot[(F−G), {z, −m, m}, {t, −m, m}, PlotPoints −> p]
(* Electric (or Magnetic) Energy *)
    ContourPlot[(F−G)², {z, −m, m}, {t, −m, m}, PlotPoints −> p]
(* Current (or Voltage) *)
    ContourPlot[(F+G), {z, −m, m}, {t, −m, m}, PlotPoints −> p]
(* Magnetic (or Electric) Energy *)
    ContourPlot[(F+G)², {z, −m, m}, {t, −m, m}, PlotPoints −> p]
(* Impedance *)
    ContourPlot[(F−G)/(F+G), {z, −m, m}, {t, −m, m}, PlotPoints −> p]
(* Power *)
    ContourPlot[(F−G)*(F+G), {z, −m, m}, {t, −m, m}, PlotPoints −> p]
(* Energy *)
    ContourPlot[(F+G)²+(F−G)², {z, −m, m}, {t, −m, m}, PlotPoints −> p]
```

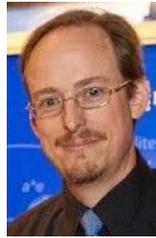

**Hans Gregory Schantz** (M'95–SM'04) was born in Detroit, Michigan, on September 27, 1966. He received degrees in engineering and physics from Purdue University in 1988 and 1990, respectively. He earned his Ph. D. degree in theoretical physics from the University of Texas at Austin in 1995.

His work experience includes stints with IBM, the Lawrence Livermore National Laboratory, ITT Technical Institute, and the ElectroScience Lab of the Ohio State University. In 1999, he joined Time Domain Corporation in Huntsville, AL as an Antenna Engineer. His pioneering work in ultrawideband (UWB) antennas led to one of the first commercially deployed UWB antennas and the first frequency-notched UWB antennas. He is the author of *The Art and Science of Ultrawideband Antennas* (Norwood, MA: Artech House, 2005), and an inventor on a total of 40 U.S. patents. Since 2002 he has been Chief Technical Officer for Q-Track Corporation (www.q-track.com) in Huntsville, AL. His research interests include electromagnetic energy flow, UWB antennas, electrically small antennas, and low-frequency, near-field systems and propagation.

Dr. Schantz organized the Huntsville Joint Chapter of the Communications, APS, and MTT Societies in 2002. He is the current chair of this Joint Chapter. The Huntsville Section of the IEEE recognized him for significant contributions in 2006. He is also a member of the Institute of Navigation.